\documentclass[aps,prl,twocolumn,
superscriptaddress,
 amsmath,amssymb,
]{revtex4-1}

\usepackage{graphicx}% Include figure files
\usepackage{dcolumn}% Align table columns on decimal point
\usepackage{bm}% bold math
\usepackage{epstopdf}
\usepackage[colorlinks,linkcolor=red,anchorcolor=blue,citecolor=blue]{hyperref}

\begin{document}

%\preprint{Physical Review Letters}

\title{n-type Rashba Spin Splitting in a Bilayer Inorganic Halide Perovskite with External Electric Field}% Force line breaks with \\

\author{Xue\text{-}Jiao Chen}
\affiliation{
 State Key Laboratory of Luminescence and Applications, Changchun Institute of Optics, Fine Mechanics and Physics, Chinese Academy of Sciences, No.3888 Dongnanhu Road, Changchun 130033, People's Republic of China
}
\affiliation{
 University of Chinese Academy of Sciences, Beijing 100049, People's Republic of China
}
\author{Lei Liu}
 \email{liulei@ciomp.ac.cn}
\affiliation{
 State Key Laboratory of Luminescence and Applications, Changchun Institute of Optics, Fine Mechanics and Physics, Chinese Academy of Sciences, No.3888 Dongnanhu Road, Changchun 130033, People's Republic of China
}
\author{De\text{-}Zhen Shen}
 \email{dzshen@ciomp.ac.cn}
\affiliation{
 State Key Laboratory of Luminescence and Applications, Changchun Institute of Optics, Fine Mechanics and Physics, Chinese Academy of Sciences, No.3888 Dongnanhu Road, Changchun 130033, People's Republic of China
}
\date{\today}% It is always \today, today,

\begin{abstract}
In this letter, we investigated the Rashba effect of the CsPbBr$_3$ bilayers under the external electric field (EEF), with the first-principles calculations. For the PbBr$_2$ terminated bilayer, we found that only electrons experience the Rashba splitting under EEF, while holes do not. Such n-type Rashba effect can be ascribed to the surface relaxation effect that reverse the position of the top valence bands. The n-type Rashba parameter can be tuned monotonically to the maximum of 0.88 eV \AA \ at EEF of 1.35 V/nm at which the sequence of top valence bands recover to the bulk style. During this process the p-type spins will not survive in this 2D CsPbBr$_3$, that indeed hints a new way for making advanced functional spintronic devices.

\iffalse
\begin{description}
\item[Usage]
Secondary publications and information retrieval purposes.
\item[PACS numbers]
May be entered using the \verb+\pacs{#1}+ command.
\item[Structure]
You may use the \texttt{description} environment to structure your abstract;
use the optional argument of the \verb+\item+ command to give the category of each item.
\end{description}
\fi

\end{abstract}

\pacs{Valid PACS appear here}% PACS, the Physics and Astronomy
%   BlochlKresse02Monkhorst                           %display desired

\maketitle

%\tableofcontents

%\section{\label{sec:level1}Introduction}
Recently, the halide perovskites have aroused great research interests in different areas, such as photovoltaics, optoelectronics, or even catalysis and electrocatalysis\cite{Science1}, and so on. That can be ascribed to their intrinsic merits, including proper bandgap, strong optical absorption, balanced carrier mobility, long carrier diffusion length, low exciton binding energy, and most of all defect tolerance\cite{Wolf2014,Wehrenfennig2014,Dong2015,Miyata2015,Yin2014,Science2}. While defect tolerance means these perovskites would be electronically inert from alien electronic states, that will certainly give them natural advantages in many device applications, where defects or interfaces are inevitable.  Although being nonmagetic, the halide perovskites can be fabricated into spintronic devices as well, in case enough spins can be produced and well controlled.

For those nonmagnetic semiconductors as perovskites, the spin degeneracy of their electronic bands can be removed by the with inverse-symmetry breaking operations, such as the dresselhaus effect\cite{Dresselhaus1955} typically in their bulk phases, or by the Rashba effect normally around their surfaces or heterostructure interfaces \cite{Bychkov1984, Rashba1960, Grundler2000, Ganichev2004, Averkiev2006, Meier2007, Dil2008, Ishizaka2011, Sante2013, Shanavas2014, Zhang2014}. So far, the Rashba effect has been studied in many materials, including InAs quantum wells (QWs)\cite{Grundler2000,Ganichev2004}, Al$_x$Ga$_{1-x}$As QWs\cite{Averkiev2006}, bulk BiTeI\cite{Ishizaka2011} and GeTe\cite{Sante2013}, together with oxide perovskites\cite{Santander2014,Shanavas2014}. However, for halide perovskites, most of the works on Rashba splitting were on their organic-inorganic hybrid members\cite{Motta2015, Kepenekian2015, Zheng2015, Leppert2016, Hutter2016, Etienne2016, Niesner2016, Moser2016,Zhai2017}. While these halides suffer the stability problem\cite{Wang2016}, obviously it would be more meaningful to study the Rashba splitting on their more robust members, $i.e.$  the all-inorganic halide perovskites.

Experimentally, the Rashba effect of all-inorganic halide perovskites has been observed by Maya $et\ al.$ on CsPbBr$_3$ nanocrystals based on the magneto-optical measurements at cryogenic temperatures\cite{Isarov2017}.
But, for the real spintronic applications, its layer structures under electrical field would be of more interest to investigate. Indeed, by different groups\cite{Akkerman2016,Bekenstein2015}, the 2D layer structures of the CsPbBr$_3$ has been successfully synthesized from monolayer to several layers. However, for these layered halide perovskites, their Rashba performance under EEF has not been identified so far.

    \begin{figure}
      \centering
      \includegraphics[width=20em]{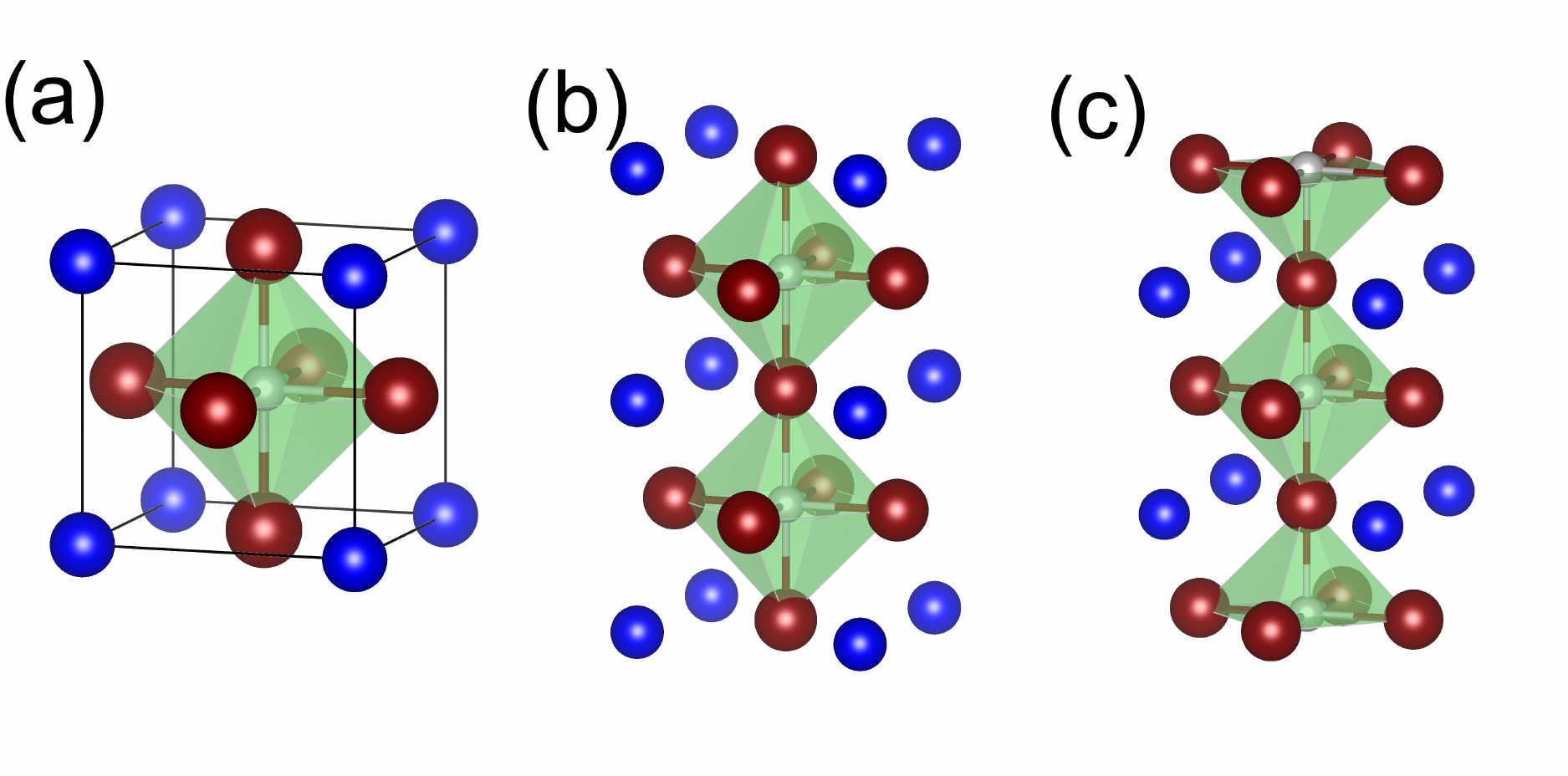}\\
      \caption{Crystal structures of CsPbBr$_3$ (a), CsBr-terminated (b) and PbBr$_2$-terminated (c) bilayers. (The blue balls represent Cs atoms, the gray balls represent Pb atoms while the brown balls represent Br atoms.)}\label{fig:fig1}
    \end{figure}

Here, with the relativistic first-principles density-functional theory (DFT) calculations, we studied the EEF-induced Rashba effect of the CsPbBr$_3$ bilayers terminated either with PbBr$_2$ or CsBr surfaces, as illustrated in FIG.\ \ref{fig:fig1}. Two types of bilayers were found exhibiting distinct spin-splitting behaviors under EEF, $i.e.$ the PbBr$_2$ one is a good Rashba material while the other is not.

%\section{\label{sec:level2}Computation Method}

The DFT calculations on CsPbBr$_3$ bilayers were performed with the projector-augmented wave\cite{Kresse03,Blochl} pseudopotentials and the Perdew-Burke-Ernzerhol\cite{Perdew} functional as implemented in the Vienna Ab-initio Simulation Package\cite{Kresse01,Kresse02}.
The experimental lattice parameters of the cubic phase CsPbBr$_3$ ($a$=5.874 \AA) \cite{Moller1958}, as shown in FIG.¡¢ \ref{fig:fig1}a, were used to construct both PbBr$_2$ and CsBr terminated slabs, where the vacuum region was set as 15  \AA.
By setting the EEF normal to their surfaces, the atomic positions of both slabs were relaxed until their residual forces were less than 0.01 eV/\AA.
Here the electron wave function was expanded using plan waves with a kinetic-energy cutoff of 300 eV. K points were generated using the Monkhorst-Pack scheme with a mesh size of 6$\times{6}$$\times{1}$.

%\section{\label{sec:level3}Results and Discussions}

    \begin{figure}%[b]
      \centering
      %\includegraphics[scale=0.35]{pic/figure2a.pdf}\\
      \includegraphics[width=20em]{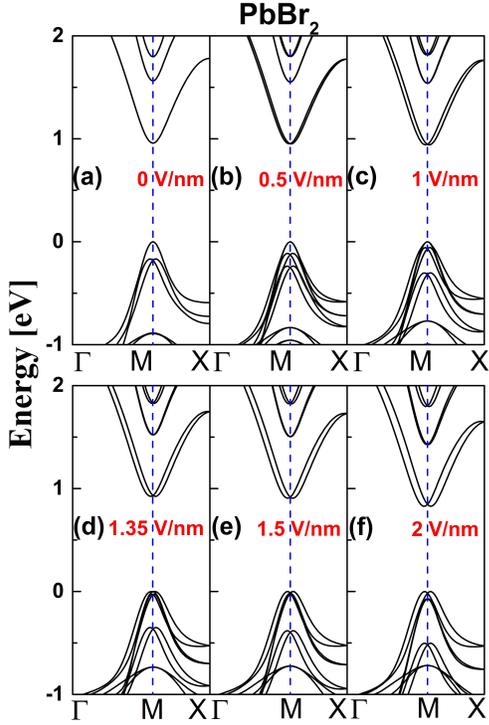}\\
      \caption{The calculated electronic band structures of the PbBr$_2$-terminated bilayer under different EEF.}\label{fig:fig2}
    \end{figure}

CsPbBr$_3$ in bulk, as shown in FIG.\ \ref{fig:fig1}a, has the symmetry of PM-3M. If cleaved along plane (001), its bilayer structures with thickness of 2$a$ can be terminated with either CsBr or PbBr$_2$ atom layers, as shown in FIG.\ \ref{fig:fig1}b and FIG.\ \ref{fig:fig1}c respectively. While the CsBr-terminated bilayers are stacked with two identical intact PbBr$_6$ octahedron layers, the PbBr$_2$ ones are linked by the units of one center PbBr$_6$ octahedron and two half octahedra symmetrically in chain. Both  PbBr$_2$ and CsBr bilayers still have the inversion symmetry, and they are nonpolar and charge neutral. As the EEF applied perpendicular to their surfaces, the inversion symmetry can be switched off accordingly and that brings the Rashba effect.

Approximately, the Rashba effect can be represented by the hamiltonian\cite{Bychkov1984,Ishizaka2011}:
 \begin{equation}
 H_R=\lambda\vec{\sigma}\cdot(\vec{E}_{z}\times \vec{k}),
 \end{equation}
where $\vec{\sigma}$ is the vector of the Pauli matrices, $\vec{E}_{z}$ is the electric field along the z axis normal to the slab surfaces, $\vec{k}$ is the electron momentum, and $\lambda$ is the coupling constant.
With this spin-orbit coupling (SOC) term, the parabolic spin-degenerate band can split into two bands as
$E^{\pm}(\vec{k})=(\hbar^2\vec{k}^2/2m^{\ast})\pm\alpha_R|\vec{k}|$, where $m^{\ast}$ is the effective mass of electron or hole and $\alpha_R$ is the Rashba parameter. Therefore, around the band extremum points, such as conduction band minimum (CBM) or valance band maximum (VBM), the Rashba parameter can be calculated as:
 \begin{equation}
 \alpha_R=\frac{2\Delta E}{\Delta k},
 \end{equation}
where $\Delta k$ is the momentum offset and $\Delta E$ is the band extremum offset respectively.

    \begin{figure}
      \centering
      \includegraphics[width=18em]{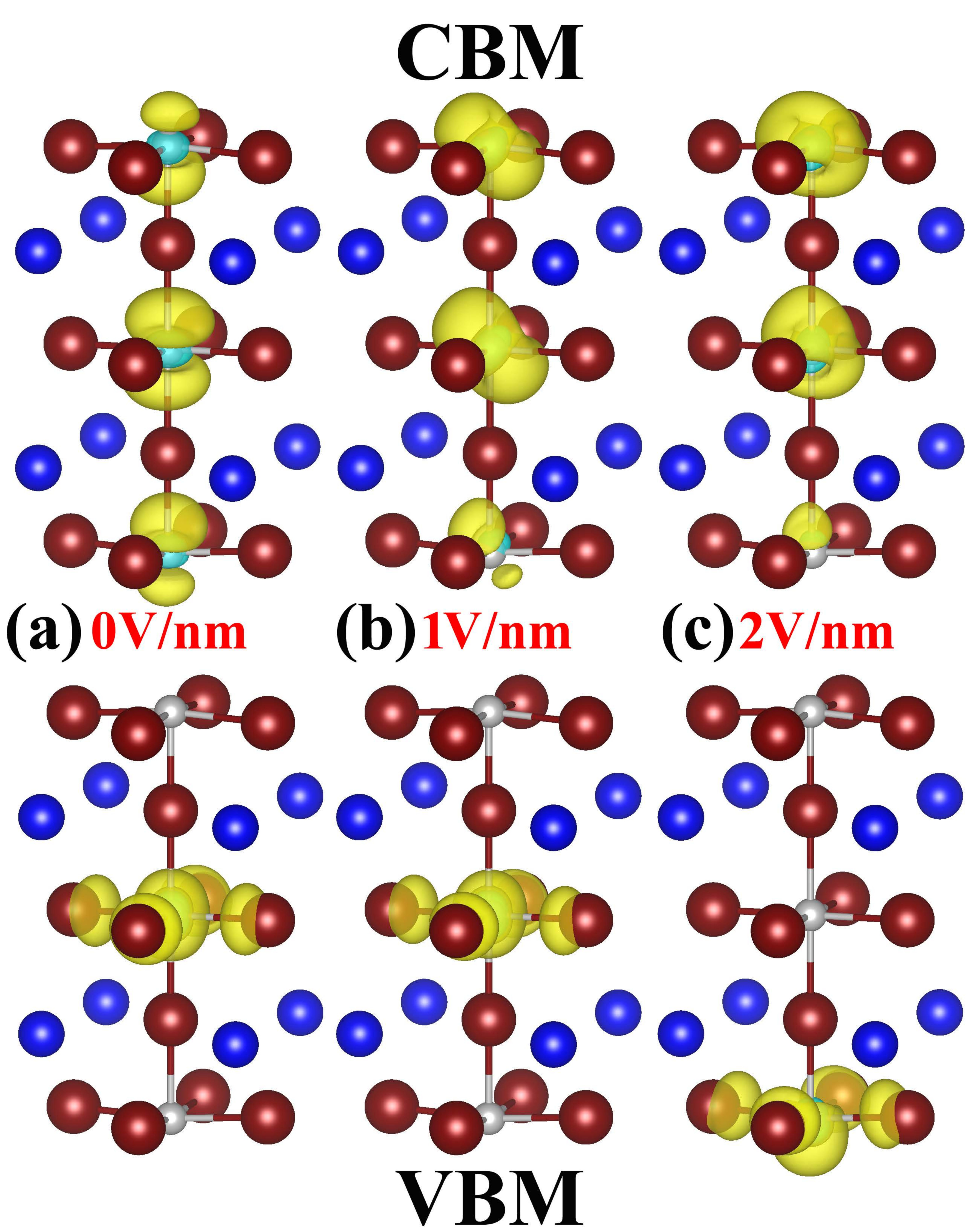}\\
      \caption{The partial charge densities of CBM and VBM the PbBr$_2$-terminated bilayer under EEF of 0, 1, and 2 V/nm.
      }\label{fig:fig3}
    \end{figure}

FIG.\ \ref{fig:fig2} presents the electronic bands around VBM and CBM with different EEFs for the PbBr$_2$ terminated bilayer. Obviously, its valence and the conduction bands behave differently upon EEF. Without EEF, both valance and conduction bands show the parabolic profiles around CBM and VBM repectively at M-point, as shown in FIG.\ \ref{fig:fig2}a. Once EEF turned on, these double-degenerate bands will split more or less as shown in FIG.\ \ref{fig:fig2}(b-f).
The lowest conduction band splits into two similar parabolic ones, with their new CBMs move away gradually from M-point to points $\Gamma$ and X separately. In comparison, after splitting the top valence ones still share the VBM at M-point, where their profiles show different curvatures as shown in FIG.\ \ref{fig:fig2}(b) and FIG.\ \ref{fig:fig2}(c). That results in the holes possess different effective masses. Moreover, with EEF, two bands just below VBM split vertically with one branch moving up and the other going down. As EEF is larger than 1.35 V/nm, the upper branch gets higher than the original top valence band. As this branch consists of two parabolic bands crossing each other at point M, that makes new VBM staying away from M-point.

    \begin{figure}
      \centering
      \includegraphics[width=25em]{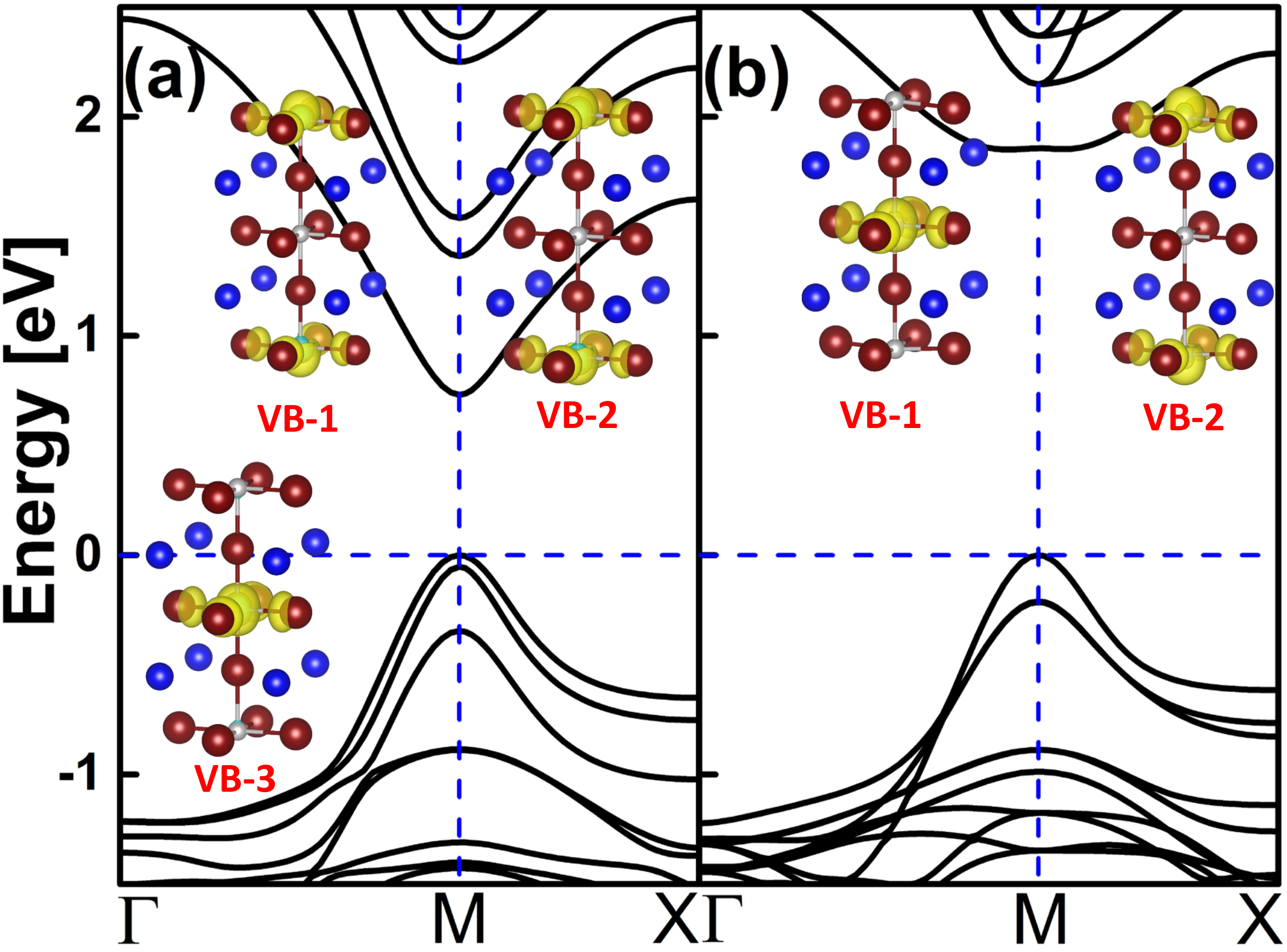}\\
      \caption{The calculated electronic band structures of the PbBr$_2$-terminated bilayer, together with the partial charge densities of top valence bands( top 1:VB-1, top 2:VB-2, top 3: VB-3) at point M, in cases of (a): without surface relaxation and with SOC; and (b): with surface relaxation and without SOC.}\label{fig:fig4}
    \end{figure}

To characterize the nature of these bands, we plotted their charge density distribution of CBM and VBM  at three selected EEF values of 0, 1 and 2 V/nm, as shown in FIG.\ \ref{fig:fig3}. Obviously, the CBM band comes from the extended p-orbitals of all Pb atoms. As shown in FIG.\ \ref{fig:fig3}a, the CBM states show the reflection symmetry about the central atomic layer. With EEF, the CBM charges becomes asymmetric, and their shapes on each Pb atom change from dumbbell-shape to spherical-like. In comparison, the initial VBM states, bonded with s-orbits of Pb and p-orbits of Br, distribute only on the central atomic layer of PbCs$_4$. As shielded by the surface electrons, such highly symmetric single-layer VBM states are not sensitive to EEF as shown in FIG.\ \ref{fig:fig3}b. Therefore, the top valence states, $i.e$ holes, don't play Rashba splitting, unless the original VBM band is surpassed by the upward surface bands as shown in FIG.\ \ref{fig:fig3}c.

The Rashba parameter of this 2D bilayer halide perovskite seems unusual, since for those 3D organic-inorganic hybrid metal halide perovskites it has been demonstrated for both electrons and holes don't show too much difference in their Rashba parameters\cite{Kim2014}. For other materials of BiTeI\cite{Ishizaka2011}, GeTe\cite{Sante2013}, black phosphorus\cite{Popovi2015}, and  organic-inorganic halide perovskites\cite{Zheng2015,Etienne2016}, their Rashba splitting always happens for both CBM and VBM bands. This asymmetric Rashba effect can be ascribed to the reversion transition of the upper valence bands of this 2D CsPbBr$_3$ bilayer. FIG.\ \ref{fig:fig4}a presents the electronic bands for the PbBr$_2$ terminated bilayer in its bulk atom positions, together with charge densities of its top three double-degenerate valence bands at point-M. Without surface relaxation, the VBM state actually comes from the surface states rather than the central layer one as presented in FIG.\ \ref{fig:fig2}a. As expected, it is SOC that removes the double degeneracy of the top valence bands of  the relaxed bilayer, as shown in FIG.\ \ref{fig:fig4}b.

\begin{figure}
      \centering
      \includegraphics[width=20em]{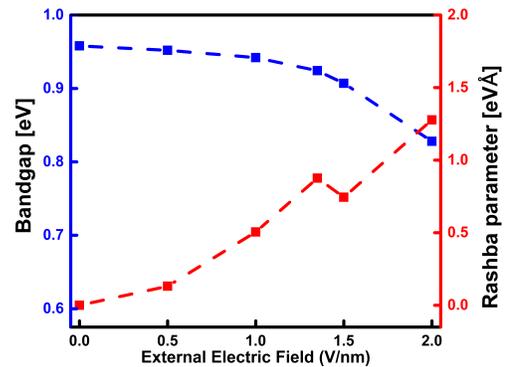}\\
      \caption{ The calculated bandgaps and Rashba parameters of CBM with a series of EEF (blue: bandgap; red: Rashba parameter).}\label{fig:fig5}
    \end{figure}

As a result of the reversion of top VBM bands, the nature of this bilayer bandgap changes consequently. As shown in FIG.\ \ref{fig:fig2}(a), the PbBr$_2$ terminated bilayer has the intrinsic direct bandgap initially. As EEF turned on, the bandgap becomes indirect, and goes to direct again after EEF gets over the band-interchange point, as shown in FIG.\ \ref{fig:fig2}(b-f). In FIG.\ \ref{fig:fig5}, we plotted the EEF-dependent bandgap, together with the calculated Rashba parameter from the CBM band based on equation (2). Basically, the bandgap decreases monotonically with the increasing EEF, but the gap value drops dramatically after the VBM transition point. Overall, that can be understood as the Stark effect\cite{Keeffe2002,Zhang2008,Qihang2012}.

Interestingly, the VBM-band switch also affects the dependence of the CBM Rashba parameter on EEF. Before this VBM flip, the CBM Rashba parameter increases continuously with EEF and reaches the maximum of 0.88 eV \AA \ at EEF of 1.35 V/nm. After that, the Rashba-parameter curve shows a kink, and then it increases with EEF again. Experimentally, the Rashba parameters of CsPbBr$_3$ were measured on nanocrystals by analysing their excitonic magneto-photoluminescence spectra\cite{Isarov2017}.  With the applied magnetic field from 0 to 8 Tesla, they found 0.2 and 0.05 eV \AA \ for the Rashba coefficient for electrons and holes respectively. Obviously, the Rashba effect is more tunable for CsPbBr$_3$ in the layer structure with the electric field. But this seems only true for those CsPbBr$_3$ layers with the proper surfaces. We found that the Rashba splitting is negligible with EEF for the CsPbBr$_3$ bilayer that is terminated with CsBr surface as shown in FIG.\ \ref{fig:fig1}b.

%\section{\label{sec:level4}Conclusions}
In summary, by theoretically examining the spin-polarized electronic bands of 2D CsPbBr$_3$ bilayers with the vertically applied electric field, we have identified the asymmetric Rashba splitting upon charge carrier type on the PbBr$_2$ terminated bilayer. It is surface relaxation that lowers the energy of the original two top valence bands and makes the original top three valence one the VBM band. While the updated VBM electronic state is protected by the surface charges and with high symmetry, the holes of this 2D CsPbBr$_3$ bilayers are inert to EEF untill the surface valence bands reverse back to the top at the EEF of 1.35 V/nm. The reversion of top VBM bands also makes the 2D CsPbBr$_3$ bilayer possess the indirect bandgap nature once EEF turned on and regain the direct style after the VBM transtion. In making spintronic devices, only n-type carriers in this 2D material will yield to the EEF-induced Rashba splitting, with the maximum parameter of 0.88 eV \AA \ at EEF of 1.35 V/nm. For the 2D layers with very large surface to bulk ratio, normally the surface relaxation will remarkably reduce the energy of the system, that act mainly on the top valence bands for stoichiometric materials. That may hint that the n-type Rashba effect could be universally observed on the similar 2D  perovskites with proper surfaces selected. That may give these 2D materials unique advantages in making advanced functional spin devices, such as in blocking undesired p-type spins.

\begin{acknowledgments}
L. L. acknowledges the support from the National Science Fund for Distinguished Young Scholars of China (No. 61525404).
\end{acknowledgments}

\bibliography{basename of .bib file}

\end{document}